\documentclass[preprint,
    superscriptaddress,
     amsmath,amssymb,
     aps,
     prx,
    longbibliography
]{revtex4-2}
\usepackage{graphicx}
\usepackage{dcolumn}
\usepackage{bm}
\usepackage{fixmath}
\usepackage[normalem]{ulem}
\usepackage[usenames,dvipsnames]{color}
\usepackage{amsmath, amssymb}
\usepackage{physics}

\usepackage[colorlinks,bookmarks=true,citecolor=blue,linkcolor=blue,urlcolor=blue]{hyperref}

\begin{document}


\title{Machine learning-based compression of quantum many body physics: PCA and autoencoder representation of the vertex function}

\author{Jiawei Zang}
\affiliation{Department of Physics, Columbia University, 538 W 120th Street, New York, New York 10027, USA}
\author{Matija Medvidović}
\affiliation{Department of Physics, Columbia University, 538 W 120th Street, New York, New York 10027, USA}
\affiliation{Center for Computational Quantum Physics, Flatiron Institute, 162 5th Avenue, New York, NY 10010, USA}
\author{Dominik Kiese}
\affiliation{Center for Computational Quantum Physics, Flatiron Institute, 162 5th Avenue, New York, NY 10010, USA}
\author{Domenico Di~Sante}
\affiliation{Department of Physics and Astronomy, Alma Mater Studiorum - University of Bologna, Bologna 40127, Italy}
\author{Anirvan M.~Sengupta}
\affiliation{Department of Physics and Astronomy, Rutgers University, 136 Frelinghuysen Road, Piscataway, New Jersey 08854, USA}
\affiliation{Center for Computational Quantum Physics, Flatiron Institute, 162 5th Avenue, New York, NY 10010, USA}
\affiliation{Center for Computational Mathematics, Flatiron Institute, 162 5th Avenue, New York, NY 10010, USA}
\author{Andrew J. Millis}
\affiliation{Department of Physics, Columbia University, 538 W 120th Street, New York, New York 10027, USA}
\affiliation{Center for Computational Quantum Physics, Flatiron Institute, 162 5th Avenue, New York, NY 10010, USA}

\date{\today}

\begin{abstract}
Characterizing complex many-body phases of matter has been a central question in quantum physics for decades. Numerical methods built around approximations of the renormalization group (RG) flow equations have offered reliable and systematically improvable answers to the initial question -- what simple physics drives quantum order and disorder? The flow equations are a very high dimensional set of coupled nonlinear equations whose solution is the two particle vertex function, a function of three continuous momenta  that describes particle-particle scattering and encodes much of the low energy  physics including  whether the system exhibits various forms of long ranged order. In this work, we take a simple and interpretable data-driven approach to the open question of compressing the two-particle vertex. We use \emph{principal component analysis (PCA)} and an \emph{autoencoder} neural network to derive compact, low-dimensional representations of underlying physics for the case of interacting fermions on a lattice. We quantify errors in the representations by multiple metrics and show that a simple linear PCA offers more physical insight and better out-of-distribution (zero-shot) generalization than the nominally more expressive nonlinear models. Even with a modest number of principal components ($\sim 10 - 20$), we find excellent reconstruction of vertex functions across the phase diagram. This result suggests that many other many-body functions may be similarly compressible, potentially allowing for efficient computation of observables. Finally, we identify principal component subspaces that are shared between known phases, offering new physical insight. We find that the vertex functions needed to describe the ferromagnetic state are not contained in the low rank description of the Fermi liquid state, whereas the vertex functions needed to describe antiferromagnetic and superconducting states are, suggesting that the latter two states emerge by amplification of pre-existing fluctuations in the Fermi liquid state while the onset of ferromagnetism is driven by a different process. These results can potentially be used in future RG calculations as a simple postprocessing step, enabling data-driven discoveries with no parameter tuning or costly training. 
\end{abstract}

\maketitle

\section{\label{introduction} Introduction}
\begin{figure}[htbp]
  \centering
  \includegraphics[width=14cm]{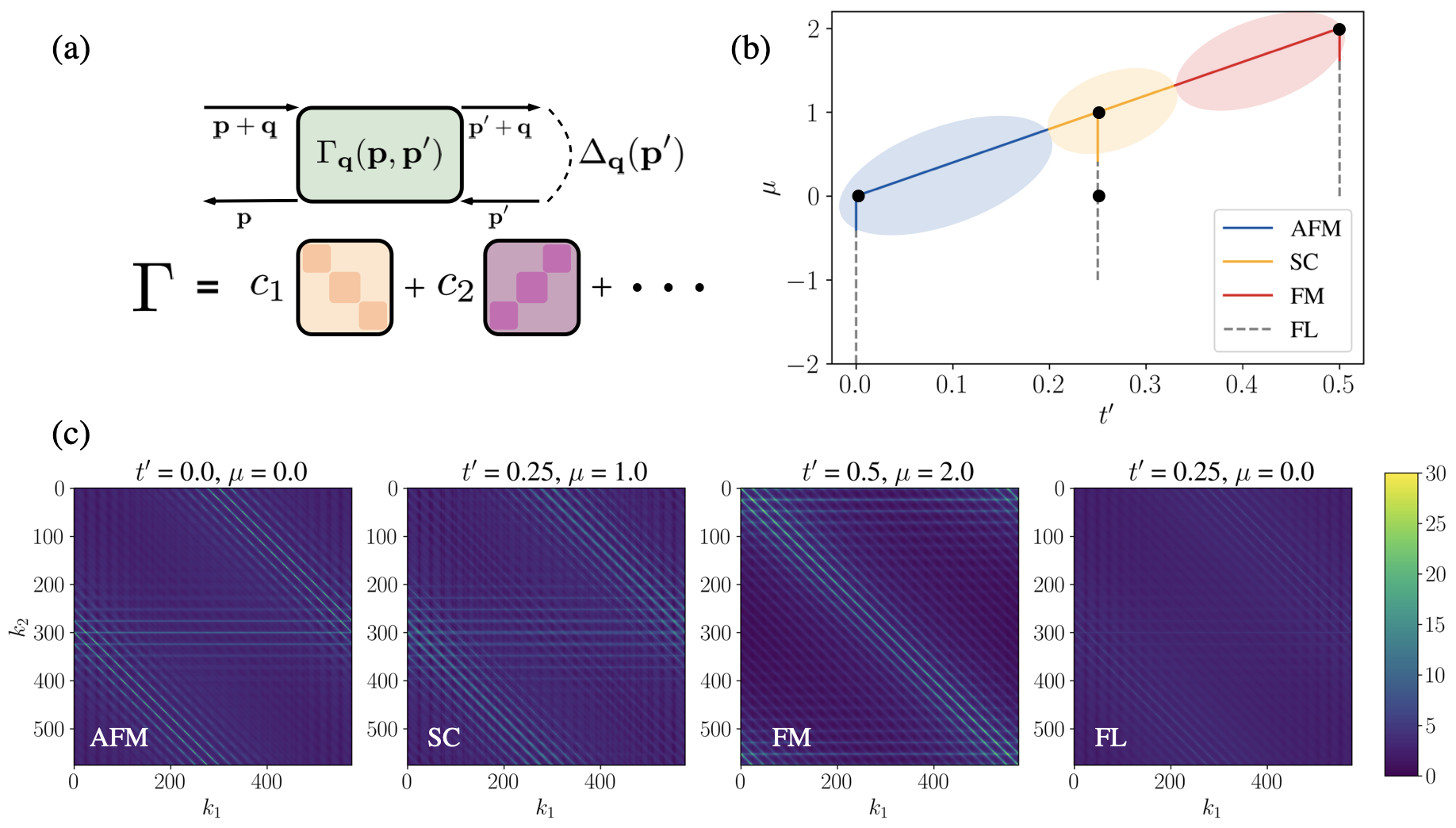}
  \caption{(a) Representation of full vertex function $\Gamma(k_1,k_2,k_3)$ scattering between incoming electrons (lines with  arrows pointing toward the vertex) and outgoing electrons (incoming holes) (lines with arrows pointing away from the vertex) of different momenta. The vertex is presented  in particle-hole notation as $\Gamma_{\bm{q}}(\bm{p}, \bm{p}')$ with $\bm{q} = \bm{k}_2 - \bm{k}_1$, $\bm{p} = \bm{k}_1$ and $\bm{p}' = \bm{k}_1 + \bm{k}_3 - \bm{k}_2$ highlighting  the scattering of a particle-hole pair with net momentum $q$ and relative momentum $2p+q$ into another pair, also of net momentum $q$ but relative momentum $2p^\prime+q$. Contracting the outgoing pair with respect to an order parameter $\Delta_{\bm{q}}(\bm{p}^\prime)$ probes tendency to particle-hole ordering with order parameter $\Delta$.  If the vertex develops long-range order at some fixed $\bm{q}$, its eigenvectors (gap functions) $\Delta_{\bm{q}}$ yield additional insights into the symmetries of the ordered phase. (b) Phase diagram of the model studied in this paper in the plane of chemical potential $\mu$ (controlling carrier concentration) and second neighbor hopping $t^\prime$ controlling the electron band structure. The shaded regions denote parameter regimes in which antiferromagnetic (blue), superconducting (yellow) and ferromagnetic (red) order occurs. The dashed lines indicate trajectories where fRG data were collected to probe different phases. (c) Momentum resolved vertex functions $\Gamma(\bm{k}_1, \bm{k}_2, \bm{k}_3)$ (full fermionic notation) at $k_3$ = 0 for different regimes of the phase diagram, indicated by black dot symbols in panel (b).}
\label{fig:fig1}
\end{figure}

One of the grand computational challenges in present-day quantum many-body physics is understanding large systems of interacting particles. Quantum physics is naturally formulated as a theory of linear operators acting on a Hilbert space whose dimension grows exponentially with the number of degrees of freedom. Exact diagonalization of any many-body Hamiltonian in this space quickly becomes unfeasible. Likewise, the infamous sign problem prevents statistically accurate solutions of generic fermionic models by means of quantum Monte Carlo methods~\cite{senechal2006theoretical} and it is generally believed that the solution of a general fermionic quantum many-body problem is NP hard \cite{Troyer05}. An alternative approach to the problem is via \emph{diagrammatic methods}, in which the two-particle vertex $\Gamma$ (called the \emph{vertex function}) plays a crucial role. The vertex function generally depends on three momenta $\bm{k}_i$ and three frequencies $\nu_i$ and can be written as $\Gamma(k_1,k_2,k_3)$ where $k_i = (\bm{k}_i, \nu_i)$ denotes a frequency-momentum four-vector. The vertex function $\Gamma$ describes the two-particle scattering from the initial states $k_2$ and $k_4 = k_1 + k_3 - k_2$ into the final states $k_1$ and $k_3$ and may also be viewed as describing the scattering of a particle-hole pair of total momentum $q=k_2-k_1$ and relative momentum $2p+q=k_1+k_2$ into another pair also of total momentum $q$ but of relative momentum $2p^\prime+q=k_3+k_4$; this latter ``particle-hole" representation is shown in Fig.~\ref{fig:fig1} (a). Knowledge of the vertex function enables insights into the many-body properties of the system, including its response to external fields as well as its tendency to develop long-range order~\cite{Abrikosov75,senechal2006theoretical, Jarrell1997, PKune2011, Toschi2012}. However, the vertex function is in general difficult to calculate and, although not exponentially large, demands large computational and memory resources with increasing system size and decreasing temperature. While there has been recent progress on compressing the frequency structure of single-particle \cite{Kaye_2022, Shinaoka_2017} and also two-particle Green's functions \cite{Shinaoka_2018,moghadas2024compressing}, low-dimensional representation of the full vertex function remains an open question \cite{Lichtenstein_2017}. Deep learning approaches have recently had success in constructing such compact latent representations of the vertex function. Authors in Ref.~\cite{DiSante2022} have used an encoder-decoder structure on top of neural ordinary differential equations~\cite{Chen2018}. A latent space is learned that can represent the entire renormalization group at a fraction of the original computational cost. This naturally raises the question: is there a concise representation of the momentum structure of $\Gamma$? 

In this paper, we use principal component analysis (PCA) and a deep convolutional autoencoder \cite{pml1Book} to address this question by compressing a large dataset of vertex functions assembled from functional renormalization group (fRG) calculations \cite{Metzner_2012} for the 2D Hubbard model \cite{Beyer_2022}. To determine the fidelity of the compressed representation we compute the pointwise mean square difference between the true and reconstructed vertices, and we compute physical quantities related to generalized susceptibilities and to the tendency of the system to order into different phases. We find in general that very low dimensional representations suffice to capture the physics accurately, suggesting the existence of a heretofore unsuspected simple underlying structure for the vertex function. We also find that PCA is markedly superior to the autoencoder in achieving the lowest possible reconstruction error for a fixed dimensionality of the representation. 

Further, quantum many-body systems may be in ordered states (for example, ferromagnetic, antiferromagnetic or superconducting) or may be in quantum disordered states such as the conventional Fermi liquid. One may ask whether the different forms of order emerge from the Fermi liquid state by an amplification of pre-existing correlations as interaction parameters are varied, or whether an order emerges via the introduction of new physics not manifest in the Fermi liquid. Our analysis enables us to address this question by quantifying the overlap of principal components of different states. Surprisingly, we find that the ferromagnetic state is not naturally contained in the Fermi liquid state, in the sense that the minimal number of principal components (4-20) that describe the Fermi liquid with good accuracy do not provide an accurate description of the ferromagnetic state vertex. On the other hand  the vertex functions appropriate to the antiferromagnetic and superconducting states can be described within the same minimal basis that describes the Fermi liquid vertices. Our findings suggest that quantum many-body physics has hitherto unsuspected structure that may allow it to be reformulated  in a compact and computationally efficient basis , and shows how comparing machine-learning-based data compressions across regimes of different physics and potentially across different systems may provide new insights.  Our findings forge a path towards refining the computation of vertex functions, offering potential advantages in computational efficiency and physics discovery.

 The rest of this paper is organized as follows. In Section~\ref{sec:model} we present the model and methods. In Section~\ref{sec:compareDLPCA} we introduce measures of the accuracy of the compression and use them to compare the reconstruction loss for PCA and autoencoder. In Section \ref{sec:FL} we study the reconstruction of ordered states using the information contained in Fermi Liquid states. In Section~\ref{sec:outlook} we give a summary, conclusion and outlook.

\section{Model and methods}\label{sec:model}
We employ the two-dimensional square lattice Hubbard model, a paradigmatic model of interacting electrons on a lattice, as a test bed for our methodology. The model may be written in standard second quantized notation using a mixed momentum ($\bm{k}$) and lattice site ($i$) representation as
\begin{equation}
    H = \sum_{\bm{k}, s} \xi_{\bm{k}} c_{\bm{k} s}^\dagger c^{\phantom{\dagger}}_{\bm{k} s}
    -\mu \sum_{i, s} n_{i s}+U \sum_{i} n_{i \uparrow} n_{i \downarrow} \,.
    \label{eq:model}
\end{equation}
Here, $c^\dagger_{\bm{k} s}$ creates an electron with momentum $\bm{k}$ and spin $s$; $\xi_{\bm{k}}=-2t[\text{cos}(k_x) + \text{cos}(k_y)]-4t^\prime \text{cos}(k_x) \text{cos}(k_y)$ describes how electrons move on the lattice ($t$ and $t^\prime$ are respectively the quantum mechanical amplitudes for electrons to move from one site to its first or second neighbor), $n_{is}=c^\dagger_{is} c^{\phantom{\dagger}}_{is}$ yields the number of electrons on site $i$, $\mu$ is the chemical potential that controls the electron density, and $U>0$ is a repulsive interaction that correlates electron motion by disfavoring configurations with two electrons per site. In this paper,  we focus on $U=3t$ with $t=1$ as our unit of energy.

For generic $\mu$ and $t^\prime$ the model is in a non-ordered \emph{Fermi liquid} state. However, for each $t^\prime$ there is a value of $\mu$ at which the Fermi surface passes through a ``van Hove" point at which the non-interacting electron density of states diverges. For chemical potentials near this point, the large density of states leads to an ordered state. Fig.~\ref{fig:fig1}(b) shows the ground state phase diagram. The solid line shows the locus of van Hove points; i.e. the $t^\prime$ dependence of the chemical potential value at which the non-interacting density of states diverges. For chemical potentials far from this line, one has a non-ordered Fermi liquid (FL) phase, and in the vicinity of the line the model exhibits either antiferromagnetic (AFM), superconducting (SC), or ferromagnetic (FM) order, depending on the value of $t^\prime$. To probe the different phases illustrated in Fig.~\ref{fig:fig1}(b), we study the evolution of the vertex function along four specific trajectories:  one follows the van Hove singularity along $\mu = 4 t'$ for $t'$ ranging from 0 to 0.5, depicted as a solid line; the other three track changes in $\mu$, from $-2 + 4t'$ to $4t'$, at fixed $t'$ values of 0.0, 0.25, and 0.5, represented by dashed lines. For each trajectory, we take 50 data points at intervals of $\Delta\mu=0.04$, resulting in a total of 200 data points.

To calculate vertex functions for the model, we employ the functional renormalization group method \cite{Beyer_2022, Honerkamp_2001}, an established computational tool for studying Fermi surface instabilities of low-dimensional interacting electron systems. At the core of the fRG method is a  set of coupled non-linear differential flow equations whose solution determines the \emph{flow} of the vertex as function of the RG scale. If momentum space is discretized into $N_{\bm{k}}$ tiles along each dimension of $d$-dimensions, $\Gamma$ is specified by $N_{\bm{k}}^{3d}$ complex numbers determined by the solution of $N_{\bm{k}}^{3d}$ coupled nonlinear equations (as common practice in fRG we only consider the lowest frequency component of the vertex). In this work, we chose $N_k=24$, such that the vertex functions extracted at the end of the fRG flow are notably high-dimensional arrays with $24^6\approx 2\times 10^8$ entries in total. To enable a comparison between PCA and the autoencoder without excessive computational costs, we down-sampled the vertex to reduce the dimension to a more manageable $144^3$, ensuring efficiency while preserving a substantial dimensionality of about  $10^6$. Details about the generation of the input data are shown in the supplemental material. Sample results, presented as two dimensional heat maps of a particular slice of the three argument vertex are shown in Fig.~\ref{fig:fig1}(c). It is noteworthy that the vertices have very substantial, nearly singular, $k$ dependence, appearing as stripes in the plots.

The calculated vertices enable one to infer the phase structure of the model via a calculation of the eigenvalues of the vertex in that particular channel where long-range order emerges. One defines an ordering eigenvector as the expectation value of a fermion particle-hole or particle particle bilinear $\langle c^\dagger_{k_1}c_{k_2}\rangle$ or $\langle c_{k_1}c_{k_2}\rangle$ (spin indices not explicitly notated) and forms an eigenvalue equation by contracting this object with appropriate indices of $\Gamma$ as shown in Fig.~\ref{fig:fig1}(a). The onset of order is signalled by a divergent eigenvalue and the bilinear with the most divergent eigenvalue $\lambda$ determines the preferred order parameter. In our fRG calculations we stop the  renormalization group flow if the magnitude of the vertex function exceeds some preset multiple of the electronic bandwidth $D$ (we choose $4D$ as the threshold) and calculate the eigenvalues and eigenvectors for different long-range orders in a post-processing step. It is important to note that in the computational basis the divergence in the eigenvalue arises not from a divergence in  a few entries of the matrix but rather from the coherent combination of a  sum over many elements, and yet the combination of elements that produces a divergent eigenvalue is identified by the dimensional reduction procedure.

 We executed PCA using a randomized SVD solver \cite{halko2011finding}. This linear method identifies the principal components, or axes, that maximize variance, thereby transforming the original data into a new coordinate system. The significance of each coordinate is ranked by the amount of variance in the data which can be captured. In effect, retaining the $M$ most significant SVD principal components defines a $M$ dimensional subspace of the $10^6$ dimensional space of all possible vertices spanned by vertices $\Gamma_{1...M}$ (each with a complicated internal structure) and the compression is the approximation that any physical vertex $\Gamma$ can be represented to sufficient accuracy as $\hat{\Gamma}=\sum_{i=1..M}a_i\Gamma_i$.

Conversely, the autoencoder is a non-linear method designed to efficiently encode a given set of data. It compresses the data into a low-dimensional latent space by applying a flexible parametrized transformation (the \textit{encoder}). A second model (the \textit{decoder}) is trained to reconstruct the original dataset as accurately as possible. The vertex is then approximated as a nonlinear function (specified by the decoder) of a number of parameters equal to the latent space dimension. We use a five-layer convolutional neural-network (CNN,~\cite{LeCun2015}) architecture for both the encoder and decoder with layer norms and GeLU activations~\cite{Hendrycks2016}. In addition to transposed convolutions, we use non-parametric upsampling after each layer.The reconstruction error, defined in Eq.~\ref{eq:epsilon_def}, serves as the cost function. Each autoencoder was trained with approximately 4 GPU-hours on a single Nvidia A100 card. 

\section{Compression loss: Comparison between PCA and the autoencoder}\label{sec:compareDLPCA}

\begin{figure}[ht]
  \centering
  \includegraphics[width=16cm]{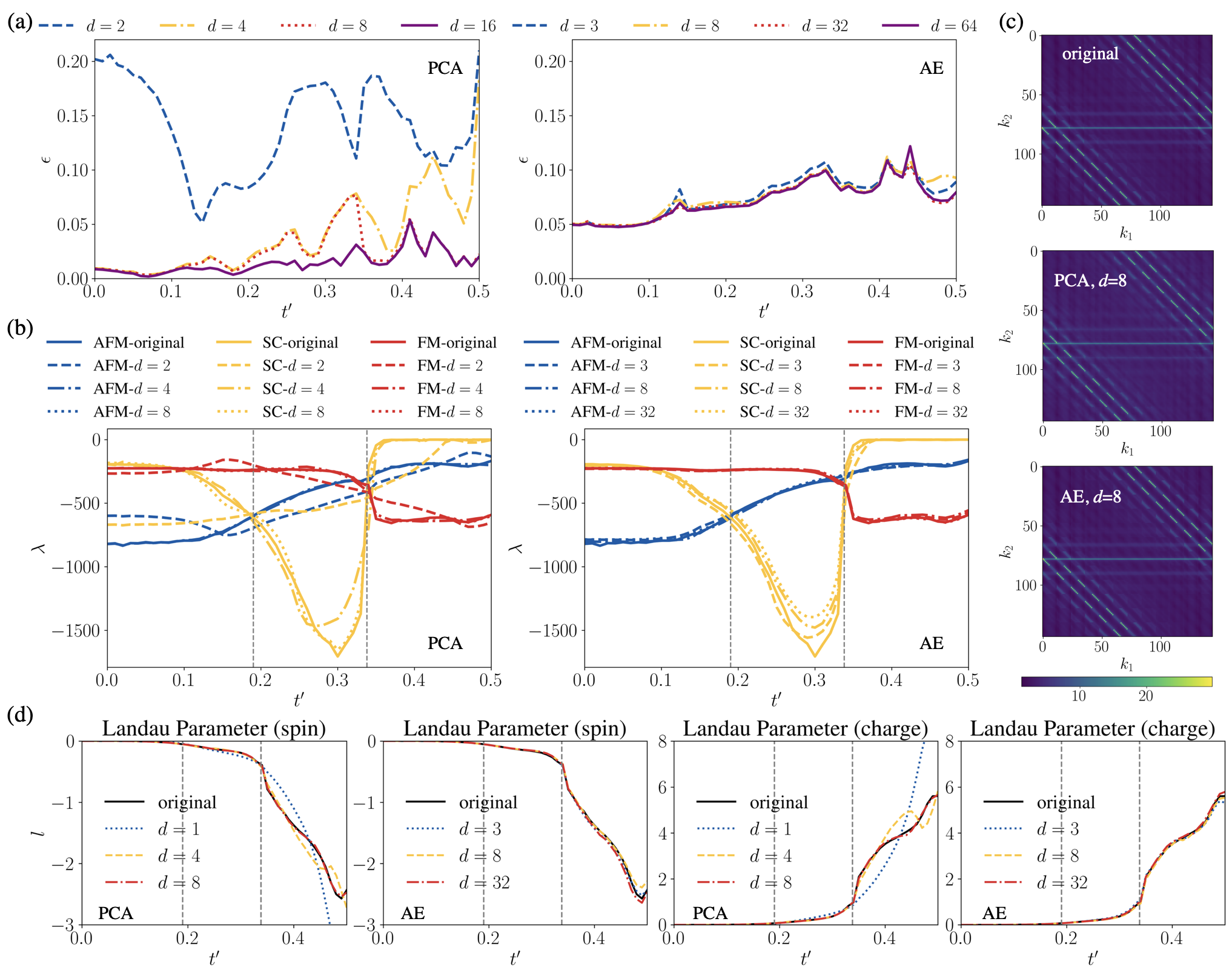}
  \caption{(a) The reconstruction error $\epsilon$ (Eq.~\ref{eq:epsilon_def}) at different $t'$ for PCA and the autoencoder (AE), each with varying latent dimension $d$. (b) The eigenvalues associated with different physical orders calculated along the van Hove trajectory. Vertical grey lines denote the boundaries between different phases as indicated by a change in the lowest-lying $\lambda$. We compare the eigenvalues calculated from the original vertex with those obtained after reconstructing it from AE and PCA at different $d$. (c) Visual comparison between the original vertex at fixed $k_3=0.0$ and $t’=0.0$  and its reconstructions. (d) Landau Parameters for spin and charge channels calculated from the original and the reconstructed vertices.}
\label{fig:compare_error}
\end{figure}
In this section we introduce measures of the accuracy of the compression. we use three error metrics: the direct (pointwise) reconstruction error of the vertex, given by 
\begin{equation}\epsilon=\frac{||\hat{\Gamma}-\Gamma||_{2}}{||\Gamma||_{2}} ,
\label{eq:epsilon_def}
\end{equation}
the reconstruction accuracy of the leading eigenvalue  $\lambda$~\cite{Zhai_gap} for several instability channels, and the Landau parameters, averages over the Fermi surface of the vertex functions weighted by angular and spin factors (see Appendix for details).  For simplicity, we restrict our study to the ordered states only, which is along the van Hove trajectory of $\mu=4t'$. But the results are similar for non-ordered states.  We use 50 vertex functions with \(t'\) ranging from $0$ to $0.5$ in increments of $\Delta t' = 0.01$ and took a random 80/20 train/test split of the input data, and subsequently evaluate the error metrics utilizing the entire dataset. To assess the quality of the compression  As shown in Fig.~\ref{fig:compare_error}(a), PCA shows decreasing reconstruction errors with increasing latent dimension $d$, whereas the autoencoder's error remains roughly constant across different $d$ values. Fig.~\ref{fig:compare_error}(b) shows the eigenvalues associated with AFM, SC, and FM order along the van Hove trajectory, with the smallest eigenvalues indicating the system's propensity towards the respective ground state. Transition points are marked by dashed vertical lines. For PCA, transitions in the reconstruction data align well with the original calculations when the latent dimension exceeds $d = 4$, and the accuracy keeps improving with higher $d$, becoming comparable to the autoendocder errors for $d\geq 4$ and clearly superior to the autoencoder errors for $d\geq 16$. In contrast, the autoencoder maintains roughly the same accuracy across various $d$. 

Fig.~\ref{fig:compare_error}(d) shows the Landau parameters $l$ for spin and charge channels. The Landau parameters \cite{Abrikosov75} are canonical quantities defined in the theory of the Fermi liquid that are related to susceptibilities  and collective mode frequencies. For PCA, the amplitude and tendency of $l$ are correctly captured even at $d=1$, indicating that the first PCA axis contains substantial information about the development of the Landau parameter, which suggests the presence of ferromagnetic states. However, the accuracy of $l$ is significantly higher for $d=8$ compared to $d=1$. For the autoencoder, the accuracy remains high for different $d$. 

Lastly, Fig.~\ref{fig:compare_error}(c) presents a side-by-side comparison of a test slice of the original vertex function \(\Gamma(\bm{k}_1, \bm{k}_2, \bm{k}_3)\) and its reconstructions at \(\bm{k}_3=0\) and \(t'=0\). We see that a relatively small number of dimensions is sufficient to represent both the vertex and physical quantities, including those relating to divergences of particular eigenvalues. As noted above, in the computational basis the divergent eigenvalues do not arise from divergences in the entries in the $\Gamma$ matrix. Our findings strongly suggest that a limited number of dimensions is sufficient to gather the essential information contained in the vertex, including information about its singularities. 

The exact reason behind the superior performance of the PCA when compared against a parameterized autoencoder neural network remains an open question. While we leave an exhaustive search over possible architecture choices for future research, our experiments with convolutional networks in three dimensions suggest that local filters fail to capture essential features in interaction vertices with a strong global sparsity structure. In contrast, performing a global linear rotation, PCA is able to capture global patterns. For other tasks on larger vertex datasets, we conjecture that using PCA as a preprocessing or input layer would result in a more expressive model.

\section{Fermi Liquid and Ordered States}\label{sec:FL}

\begin{figure}[ht]
  \centering
  \includegraphics[width=15cm]{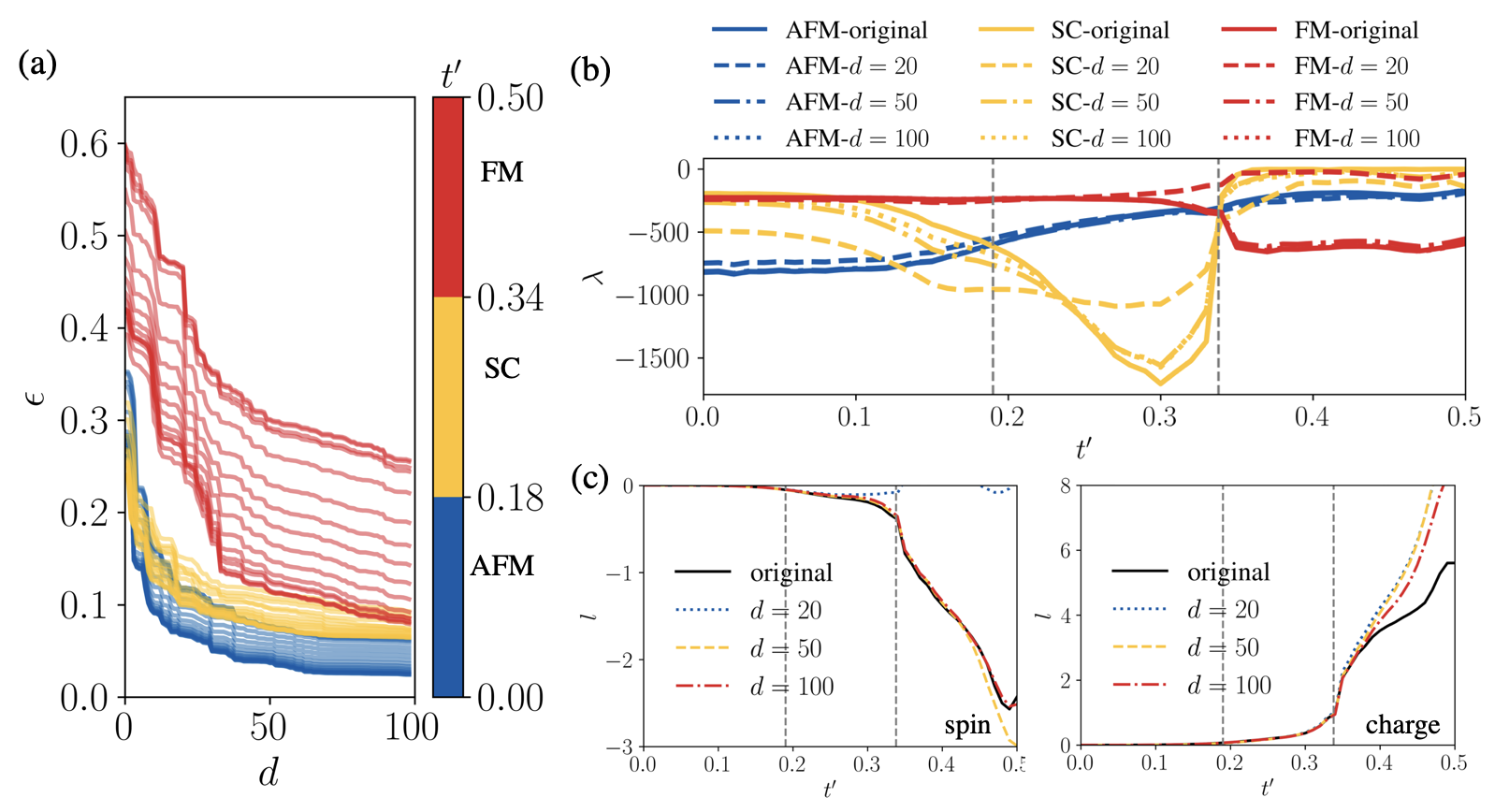}
  \caption{(a) Reconstruction error  $\epsilon$ (Eq.~\ref{eq:epsilon_def}) for all ordered states along van Hove trajectory ($t'=0-0.5$) as a function of the latent dimension number $d$ of PCA bases derived from all FL states. The data points are color-coded to indicate their association with FM, SC and AFM states. (b, c) The eigenvalues associated with different physical orders and Landau parameters calculated along the van Hove trajectory. It compares the original vertex data with the FL PCA reconstruction that utilize different number (20, 50, 100) of bases. }
\label{fig:fl_order}
\end{figure}
%

An interesting physics question concerns whether an ordered state may be understood as arising from an interaction-driven enhancement of fluctuations that are already present in the Fermi Liquid state, or whether the onset of the ordered state signals new physics not evident from an analysis of the Fermi Liquid state. The compression methods introduced here offer a new perspective on this question. We use the PCA methodology to obtain a reduced dimension subspace that accurately represents the Fermi liquid vertices (117 vertices in total). We then assess the accuracy  with which the vertices in the ordered state regions of the phase diagram may be represented within the subspace defined from the Fermi liquid phase. 

Panel (a) of Fig. \ref{fig:fl_order} shows the pointwise reconstruction error $\epsilon$ as a function of the size of $d$ of the PCA subspace used to represent the Fermi liquid vertices computed along the van Hove line for a large number of $t^\prime$ values, shown as different lines color-coded according to the relevant type of order. We see that for relatively small $d\leq 10$ the Fermi-Liquid-defined subspace provides a relatively poor compression of the ordered state vertices, but that as the subspace dimension is increased the vertices in the antiferromagentic and superconducting phases become relatively well represented while the vertices in the ferromagnetic region remain very poorly described. This analysis indicates that none of the ordered states are accurately represented by the leading terms of Fermi Liquid state, but that considering the vertex function's high dimension ($\sim10^6$), the superconducting and antiferromagnetic states are moderately well represented in terms of smaller but non-negligible contributions to the state. On the other hand, the ferromagnetic state vertex function appears to require qualitatively new components: the structure learned or inferred from the Fermi Liquid phase does not transfer well to the ferromagnetic phase.  Additionally, the errors indicate a separation at about $t'=0.34t$, coinciding with the FM and SC state transition.

Fig.~\ref{fig:fl_order}(b) presents the eigenvalues from the original and reconstructed vertices for various $d$. We see that the leading eigenvalue of the AFM phase is very well represented even at $d=20$ consistent with the lower $\epsilon$ but that for the SC vertex the representation derived from the FL phase, while qualitatively reasonable at $d=20$ only begins to become accurate by $d=50$. The FM eigenvalue is qualitatively incorrect at $d=20$ and becomes correct for $d=50$. Fig.~\ref{fig:fl_order}(c) presents Landau parameters from the original and reconstructed vertices. The Landau parameter in the spin channel initially presents inaccuracies at $d=20$, attaining precision only when $d\geq 50$, similar to the behavior observed for the FM eigenvalue. Conversely, in the charge channel, while the parameter is approximately correct starting from $d=20$, it fails to achieve high precision even at $d=100$. This observation confirms that a low dimensional representation of the Fermi liquid vertices does not transfer well to a representation of physically relevant quantities such as pairing eigenvalues and Landau parameters.

\begin{figure}[ht]
  \centering
  \includegraphics[width=15cm]{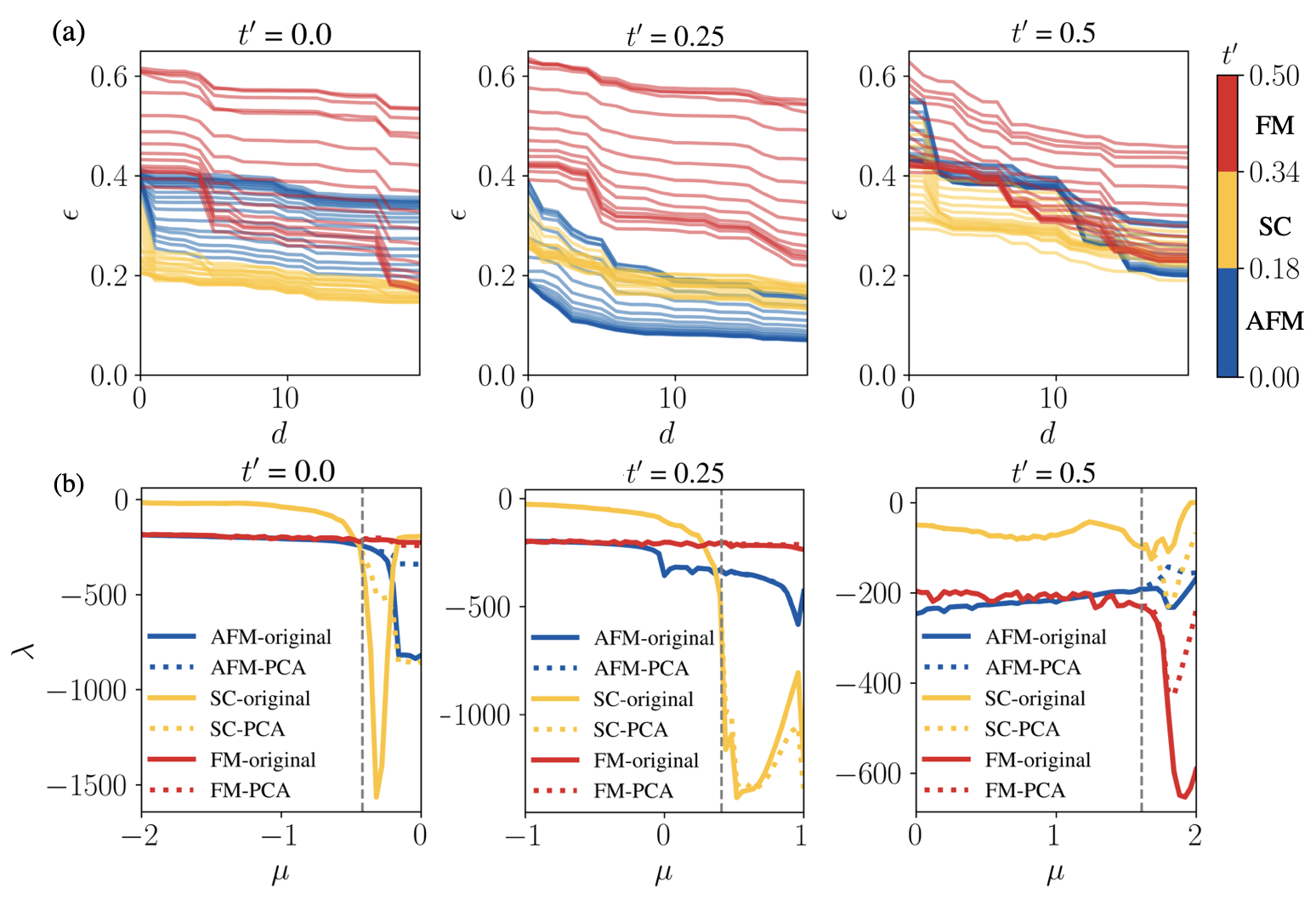}
  \caption{(a) Reconstruction error $\epsilon$ (Eq.~\ref{eq:epsilon_def}) for the ordered states along the van Hove trajectory as a function of the latent dimension $d$ of PCA bases derived from FL states at $t'$=0.0, 0.25 and 0.5. The data points are color-coded to indicate their association with FM, SC and AFM states. (b) The eigenvalues corresponding to different physical orders calculated along the Fermi liquid to ordered state transition at $t'$=0.0, 0.25, and 0.5. The vertical dashed line in each figure marks the onset of an ordered state in the fRG calculation, with the dataset on the left-hand side of these lines being employed for the PCA analysis carried out in (a). The dashed cruves represent the calculations using the reconstructed vertex with $d=20$.}
\label{fig:fl_order_t}
\end{figure}

In Fig.~\ref{fig:fl_order_t} we present a more refined analysis. We consider a sequence of Fermi Liquid states defined by varying $\mu$  within the Fermi liquid regime  at fixed $t^\prime=0.0,0.25,0.5$ (vertical lines in Fig.~\ref{fig:fig1}(b)). The three sequences of $\mu$ terminate in different ordered states, so we expect that the sequence of Fermi liquid states will be characterized by a growth of fluctuations in a given channel, leading eventually to the divergent fluctuations implying order, and that the evolution of the vertex will reflect this growth. We then define a PCA subspace for each sequence, and ask how well the ordered state vertices are described within the PCA subspace. We find that the reconstruction errors are larger than those obtained from PCA subspaces constructed from the entire Fermi liquid data set. Specifically, for bases constructed from vertices at $t'=0.0$, i.e. in the vicinity of AFM order, SC ordered states exhibit the lowest reconstruction error compared to AFM and FM orders. Similarly, for bases constructed from $t'=0.25$, which is close to superconducting order, AFM ordered states yield the smallest $\epsilon$. To develop a qualitative understanding of this circumstance, we calculate the eigenvalues of different orders along the Fermi liquid to ordered state transition as shown in Fig.\ref{fig:fl_order_t}(b). For $t'=0.0$ and $t'=0.25$, we observe, that concomitantly with the incipient order at the van Hove line ($\mu = 4t'$), i.e.~AFM order or singlet-pairing, the respective other ordering channel also shows an increase in the absolute value of $\lambda$. This indicates, that some fingerprint of fluctuations in a competing channel must be present in the vertex function, which apparently allows the PCA analysis to extract some information about the vertices in the ordered phase from the FL data. Note that this is not the case for $t'= 0.5$ (see the righmost panel of Fig.\ref{fig:fl_order_t}(b)), where the AFM and SC eigenvalues decrease in absolute magnitude when ferromagnetic order starts to set in.

\section{Outlook}\label{sec:outlook}

In this paper, we have shown that the two-particle vertex obtained from fRG calculations on the two dimensional Hubbard model, is well represented by a point in a low (i.e.~4-12) dimensional latent space derived from PCA or from  an autoencoder. We observe that the sharp structures in the exactly calculated vertices even in the Fermi liquid state show that the success of the compression does not arise simply because the vertices are smooth (and hence approximated by a few low-order polynomials), and that the successful calculation of the pairing eigenvalues means that the compression can capture singularities arising from coherent sums over many non-divergent terms in the computational basis form of the vertex. The fact that the three-fold momentum dependence of the vertex can be so strongly compressed suggests that there is a hitherto unsuspected structure in the vertex functions and perhaps more generally in the correlation functions of the Hubbard model at moderate interaction strength, and also offers the hope that much more computationally efficient schemes for calculating the vertex can be devised. Our findings thus call for further investigation in this direction. 

Note that the compression achieved with the two data-driven approaches discussed in this manuscript is even more efficient than some of the known techniques in the fRG literature. The truncated-unity (TU) approximation \cite{Beyer_2022, Lichtenstein_2017}, for example, assumes a weak-dependence of $\Gamma$ on the two fermionic momenta in the particle-particle or particle-hole momentum/frequency convention and, consequently, represents it by a set of analytical form factors. Although the number of form factors required to converge TU-fRG calculations is usually small (order $10$) \cite{Hille_2020, Gneist_2022}, fine resolution in the transfer momentum is required to resolve incipient long-range orders, resulting in many more degrees of freedom than identified in our data-driven analysis.

We also found that a low-rank representation of a Fermi liquid state was not always transferable to the ordered states, suggesting that the ordered phases do not evolve in a simple way from structures of the non-ordered Fermi liquid. The non-transferability was particular acute in the ferromagnetic phase, suggesting that this physics is simply qualitatively different from the physics of the Fermi liquid phase. One important long-term goal would be the direct calculation of vertex functions in the compressed basis that we unraveled in this manuscript. This would require the projection of the fRG or other many-body equations into the basis suggested by the PCA analysis. Further, our finding that the PCA methods provide a substantially more efficient representation than the standard autoencoder methods suggests that research into the optimizations required in the autoencoder method may be beneficial.

\begin{acknowledgments}
J.Z. and A.J.M. acknowledge support from the NSF MRSEC program through the Center for Precision-Assembled Quantum Materials (PAQM) - DMR-2011738. The research leading to these results has received funding from the European Union’s Horizon 2020 research and innovation programme under the Marie Sk{\l}odowska-Curie Grant Agreement No. 897276 (D.D.S.). M.M. was supported by the CCQ graduate fellowship in computational quantum physics. The Flatiron Institute is a division of the Simons Foundation.
\end{acknowledgments}


\nocite{*}

\bibliography{apssamp}

\end{document}


\title{Supplementary Material for Machine learning-based compression of quantum many body physics: PCA and autoencoder representation of the vertex function}

\author{Jiawei Zang}
\affiliation{Department of Physics, Columbia University, 538 W 120th Street, New York, New York 10027, USA}
\author{Matija Medvidović}
\affiliation{Department of Physics, Columbia University, 538 W 120th Street, New York, New York 10027, USA}
\affiliation{Center for Computational Quantum Physics, Flatiron Institute, 162 5th Avenue, New York, NY 10010, USA}
\author{Dominik Kiese}
\affiliation{Center for Computational Quantum Physics, Flatiron Institute, 162 5th Avenue, New York, NY 10010, USA}
\author{Domenico Di~Sante}
\affiliation{Department of Physics and Astronomy, Alma Mater Studiorum - University of Bologna, Bologna 40127, Italy}
\author{Anirvan M.~Sengupta}
\affiliation{Department of Physics and Astronomy, Rutgers University, 136 Frelinghuysen Road, Piascataway, New Jersey 08854, USA}
\affiliation{Center for Computational Quantum Physics, Flatiron Institute, 162 5th Avenue, New York, NY 10010, USA}
\author{Andrew J. Millis}
\affiliation{Department of Physics, Columbia University, 538 W 120th Street, New York, New York 10027, USA}
\affiliation{Center for Computational Quantum Physics, Flatiron Institute, 162 5th Avenue, New York, NY 10010, USA}

\maketitle

These supplementary materials contain the details of analytic calculations as well as additional numerical details supporting the results presented in the main text.

\section{Generation of the input data}

For the generation of the input data we use the \emph{Grid-fRG} method detailed in Ref.~\cite{Beyer_2022} together with the \emph{temperature flow} developed by Honerkamp and Salmhofer~\cite{Honerkamp_2001}. Within this scheme, the two-particle vertex $\Gamma(\boldsymbol{k}_1, \boldsymbol{k}_2, \boldsymbol{k}_3)$ is discretized on a uniform $N_{\boldsymbol{k}} \times N_{\boldsymbol{k}}$ Brillouin zone mesh for each $\boldsymbol{k}_i$ (here $N_{\boldsymbol{k}} = 24$) and only its zero frequency components are considered. Furthermore, self-energy corrections are neglected, such that the single-particle Green's function $G(\boldsymbol{k}, i\nu) = (i\nu + \mu - \epsilon(\boldsymbol{k}))^{-1}$ corresponds to an unrenormalized Green's function $G_0$ with chemical potential $\mu$ and $\epsilon(\boldsymbol{k})$ the dispersion for the non-interacting part of Hamiltonian. This way, frequency summations on the right-hand side of the fRG flow equations can be carried out analytically, and only the momentum integral over the Brillouin zone demands a numerical solution. However, since $G_0$ peaks sharply at the Fermi surface, especially at low temperatures, the coarse mesh used for the vertex is insufficient to perform the Brillouin zone integration with reasonable accuracy. Consequently, we refine the mesh for $G_0$ by an additional $n_{\boldsymbol{k}} \times n_{\boldsymbol{k}}$ points (this work: $n_{\boldsymbol{k}} = 64$) around each coarse mesh point. Note that this choice of grid parameters requires non-uniform quadrature weights. 

\section{Calculation of Landau parameters from fRG}

For a fully rotation invariant normal Fermi liquid, the Landau parameters are typically obtained by expanding the angular dependence of the particle-hole vertex function in Legendre polynomials for small transfer momentum $\bm{q}$ (the other two momenta are pinned to the Fermi surface) and frequency $\omega$. Notably, the two limits $\bm{q} \to 0$ and $\omega \to 0$ do not commute \cite{abrikosov_gorkov, Krien_2019}, giving rise to two distinct vertex functions describing the low-energy physics. From the fRG calculations we can only infer the $\omega / |\bm{q}| \to 0$ component ($\Gamma^k$ as opposed to $\Gamma^{\omega}$ in the notation of Ref.~\cite{abrikosov_gorkov}), i.e.~we take the limit $\omega \to 0$ first. The Legendre coefficients for $\Gamma^k$ and $\Gamma^{\omega}$, however, have a one-to-one relation \cite{abrikosov_gorkov}, which allows us to focus on the decomposition of $\Gamma^k$ (we drop the superscript in the following) here and in the main text.

Since rotation symmetry for the Hubbard model is reduced to the point group of the square lattice, we need to invoke the basis functions of the irreducible representations of $C_{4v}$, instead of Legendre polynomials. The Landau parameters in channel $\alpha \in \{\text{charge}, \text{spin} \}$ for the irrep $\mathcal{R}$ are then computed as 
\begin{align}
    l^{\alpha}_{\mathcal{R}} = \frac{1}{Z^2} \sum_{\bm{k}, \bm{k'}} \Gamma^{\alpha}(\bm{q} \approx 0, \bm{k}, \bm{k'}) \Delta(\epsilon(\bm{k}) - \mu) \Delta(\epsilon(\bm{k'}) - \mu) \phi_{\mathcal{R}}(\bm{k}) \phi_{\mathcal{R}}(\bm{k}') \,,
\end{align}
where $\Delta$ is a Gaussian distribution with zero mean and finite variance $\delta$ such that only contributions close to the Fermi level are taken into account. To gauge our calculations with respect to $\delta$, the normalization factor $Z$ is determined by
\begin{align}
    Z = \sum_{\bm{k}}  \Delta(\epsilon(\bm{k}) - \mu) \,.
\end{align}
In practice, we can only consider a finite subset of basis functions from the infinite number of available $\phi_{\mathcal{R}}$. Yet, these are naturally grouped in terms of the length scales which they capture in real space (see Ref.~\cite{Platt_thesis}) and an expansion of $\Gamma^{\alpha}$ in lattice-harmonics is thus expected to converge rather quickly (at least for the fermionic momentum dependence). We therefore consider only the nearest-neighbor contributions with basis functions
\begin{align}
    \phi_{\mathcal{A}_1}(\bm{k}) &= \cos(\bm{k}_x) + \cos(\bm{k}_y) \\
    \phi_{\mathcal{B}_1}(\bm{k}) &= \cos(\bm{k}_x) - \cos(\bm{k}_y) \\
    \phi_{\mathcal{E}, 1}(\bm{k}) &= \sin(\bm{k}_x) \\
    \phi_{\mathcal{E}, 2}(\bm{k}) &= \sin(\bm{k}_y) \,.
\end{align}
Note that $\phi_{\mathcal{A}_2 / \mathcal{B}_2} = 0$ for nearest-neighbors. Further, there is only one basis function for the one-dimensional $\mathcal{A}_i$ and $\mathcal{B}_i$ representations, whereas $\mathcal{E}$ is two-dimensional and thus requires two basis functions to be properly characterized for each real-space `shell`. For the doped $t$-$t'$ Hubbard model studied in the main text, we found that only the isotropic contribution $l_{\mathcal{A}_1}$ showed significant structure when approaching an ordered phase and we have thus limited our display of results to the Landau parameters for this particular irrep.

\bibliographystyle{apsrev}
\bibliography{apssamp}